\documentclass[journal]{IEEEtran}
\ifCLASSINFOpdf
   \usepackage[pdftex]{graphicx}
\else
\fi
\usepackage{amsmath}

\usepackage{breqn}
\usepackage{xcolor}
\usepackage{graphicx} 
\usepackage{caption,subcaption}
\usepackage{mathtools}
\usepackage{booktabs}
\usepackage{algorithm}
\usepackage{algpseudocode}
\usepackage{multirow}
\usepackage{array}

\usepackage{mathtools}
\usepackage{amsfonts}
\usepackage{booktabs}
\usepackage{algorithm}
\usepackage{algpseudocode}
\usepackage{amsmath}

\begin{document}
%\title{Siamese Autoencoder Designed for Small-scale Subject-independent EEG Analysis}

%\title{Trend Descriptor and Siamese Autoencoder Designed for Small-scale Subject-independent EEG-based Tinnitus Neurofeedback}
\title{Disentangled and Side-aware Unsupervised Domain Adaptation for Cross-dataset Subjective Tinnitus Diagnosis}
%\title{A Novel Machine-learning tool Diagnoses Hidden Hearing Loss in Listeners with Subjective Tinnitus}


\author{Yun Li, Zhe Liu, Lina Yao, \IEEEmembership{Senior Member, IEEE}, Jessica~J.M.Monaghan,~\IEEEmembership{Member,~IEEE}, and~David~McAlpine
\thanks{Y. Li, Z. Liu, and L. Yao are with the School of Computer Science and Engineering, University of New South Wales, Sydney, NSW 2052, Australia (e-mail:  zheliu912@gmail.com;
yun.li5@unsw.edu.au;
lina.yao@unsw.edu.au).}
\thanks{J. J.M.Monaghan is with National Acoustic Laboratories, The Australian Hearing Hub, Sydney, NSW, 2109, Australia
Macquarie University, The Australian Hearing Hub, NSW, 2109, Sydney, Australia. Australia e-mail: jessica.monaghan@gmail.com.}
\thanks{D. McAlpine is with Department of Linguistics, Macquarie University, NSW, 2109, Sydney, Australia. Australia e-mail: david.mcalpine@mq.edu.au}}


%Telephone: (800) 555--1212, Fax: (888) 555--1212}


% \markboth{Journal of \LaTeX\ Class Files,~Vol.~14, No.~8, August~2015}%
% {Shell \MakeLowercase{\textit{et al.}}: Bare Demo of IEEEtran.cls for IEEE Journals}

\maketitle

\begin{abstract}
EEG-based tinnitus classification is a valuable tool for tinnitus diagnosis, research, and treatments. Most current works are limited to a single dataset where data patterns are similar. But EEG signals are highly non-stationary, resulting in model's poor generalization to new users, sessions or datasets. Thus, designing a model that can generalize to new datasets is beneficial and indispensable. To mitigate distribution discrepancy across datasets, we propose to achieve Disentangled and Side-aware Unsupervised Domain Adaptation (DSUDA) for cross-dataset tinnitus diagnosis. A disentangled auto-encoder is developed to decouple class-irrelevant information from the EEG signals to improve the classifying ability. The side-aware unsupervised domain adaptation module adapts the class-irrelevant information as domain variance to a new dataset and excludes the variance to obtain the class-distill features for the new dataset classification. It also align signals of left and right ears to overcome inherent EEG pattern difference. We compare DSUDA with state-of-the-art methods, and our model achieves significant improvements over competitors regarding comprehensive evaluation criteria. The results demonstrate our model can successfully generalize to a new dataset and effectively diagnose tinnitus.

\end{abstract}
\begin{IEEEkeywords}
EEG signals, tinnitus diagnosis, disentangled representation, unsupervised domain adaptation
\end{IEEEkeywords}

% \fi
\IEEEpeerreviewmaketitle

\section{Introduction}

Subjective tinnitus is the perception of sounds without real acoustic stimulus and can only be heard by the patients themselves~\cite{henry2008using,reavis2012temporary}. People with tinnitus may suffer from sleep problems, difficulty concentrating, and negative emotions that may progress to psychiatric disorders~\cite{nova2006tinitus, espinosa2014tratamiento, eggermont2004neuroscience}. Subjective tinnitus is a worldwide problem, affecting around 15\% of the population~\cite{heller2003classification,nondahl2010ten}.  Therefore, it is essential to study the diagnoses and treatments of subjective tinnitus. 

Recently, there have been substantial achievements in tinnitus diagnosis. Among them, methods based on electroencephalography (EEG) signals attract much attention because 1) many works prove that there exists a strong correlation between abnormal EEG and tinnitus~\cite{eggermont2015maladaptive,ibarra2020acoustic}; 2) EEG signals have objective evaluation compared with traditional methods that are based on the Visual Analog Scale (VAS) or questionnaires~\cite{haefeli2006pain,kersten2014pain,powers2015acoustic,adamchic2014reversing}. With the development of information technology, many efforts are made to utilize machine learning and deep learning algorithms to analyze EEG signals and classify patients' tinnitus status~\cite{sun2019multi,liu2021generalizable,li2016svm,saeidi2021neural,allgaier2021deep,zhou2021tinnitus}. For example, Liu et al.~\cite{liu2021generalizable} design a neural network with 
Siamese structure to augment training EEG data. 

Despite the success of these methods, few of them equip with generalization abilities, especially when applying the models to datasets where distributions are mismatched with the training datasets. The collection and annotation of EEG signals of tinnitus are labor-intensive and time-consuming, which makes labeled training data hard to acquire. However, a large amount of labeled data are fundamental to the success of deep learning models.  Therefore,  we hope to adapt models trained on labeled datasets to other unlabeled datasets. However, EEG signals of different datasets can be sufficiently different due to EEG signals' high sensitivity to non-physiological factors, e.g., biological or physical conditions of brains~\cite{ibarra2020acoustic} and experimental settings such as time duration, sampling frequency, electrode placement, and equipment. The distribution discrepancy prevents the model's cross-dataset tinnitus diagnosis. In other EEG-based fields, a few works explore unsupervised domain adaptation or transfer learning to overcome the distribution mismatch across datasets~\cite{chai2016unsupervised,cimtay2020investigating,xu2020cross,rayatdoost2018cross}. However, in tinnitus diagnosis, methods that can generalize to other datasets without labeled data are under-explored. 

In addition, although unsupervised domain adaptation can relieve the distribution discrepancy and enable a cross-dataset diagnosis to some extent, the adaptation process inevitably causes loss of useful information and introduces noise that is irrelevant to or even harms tinnitus diagnosis. Also, the original EEG signals have low signal-to-noise ratio, i.e., the data contains noise even without adaptation to other datasets. Therefore, designing models that can perform adaptation to transfer knowledge across datasets and exclude tinnitus-irrelevant information is fundamental to improving the cross-dataset diagnosis performance.

Apart from noise caused during signal collection and domain adaptation, the side information of the ears, i.e., which ear the evoking acoustic signal is presented to, may also impair the classification. The right and left ears display different EEG patterns even they have the same condition as tinnitus or normal~\cite{mikkelsen2015eeg,lindenberg2016neural}. This derives from functional and anatomical inter-hemispheric differences in auditory cortex~\cite{shtyrov1999noise,rodrigo2008language,penhune1996interhemispheric,dorsaint2006asymmetries}. However, this is not noise that can be excluded. Thus, it is essential to adapt the information and patterns of the left and right ears at a subject level to eliminate the side influence.

We propose a novel model to perform Disentangled and Side-aware Unsupervised Domain Adaptation (DSUDA) for cross-dataset subjective tinnitus diagnosis to address the aforementioned problems. We propose Disentangled Auto-Encoder (DAE) to decouple tinnitus-related and tinnitus-unrelated information. The performance can be improved as noise is excluded and only pure information is fed into the classifier. In addition, we present the Side-aware Unsupervised Domain Adaptation (SUDA) to perform subject- and domain-level adaptation. The side-aware adaptation aligns features of left and right ears to avoid intrinsic left-right EEG differences harming the classification. The domain-level adaptation cooperates with DAE to extract domain variance, i.e., the tinnitus-unrelated information that must be excluded, and adapt the variance across domains. Then unsupervised domain adaptation can be achieved that the classifier trained on the source domain can be applied in the target domain without labeled data of the source domain. This is essential for EEG-based tinnitus diagnosis as it may reduce the high demand for labeled EEG signals, save cost, and find universal models shared across subjects, datasets, and different experimental settings.

In summary, we make three-fold contributions as follows:

\begin{itemize}
    \item We propose a novel model aiming at disentangled and side-aware unsupervised domain adaptation (DSUDA) for cross-dataset subjective tinnitus diagnosis. DSUDA combines disentangled auto-encoder and side-aware unsupervised domain adaptation, thus retaining the discrimination power when applied to the target domain.
    \item We design an adversarial training framework for the cooperation of the proposed disentangled auto-encoder and side-aware unsupervised domain adaptation. Training the two modules in an adversarial manner can distill class-related information and align data from different sides and domains.
    \item We conduct extensive experiments and ablation studies to evaluate and analyze our DSUDA. Our model yields 8.6\%, 10\%, and 7.5\% improvements over state-of-the-art (SOTA) methods in both-ears, left-ear, and right-ear accuracy, and consistently outperforms SOTA in other criteria, e.g., Negative-F1 score, Positive-F1 score, and Negative Predictive Value, etc. The results demonstrate that our model's generalization ability on new domain.
\end{itemize}

\section{Related work}
\subsection{EEG-based Tinnitus Diagnosis and Treatment}
Subjective tinnitus is defined as the perception of sound that can only be perceived by patients in the absence of an external sound source ~\cite{henry2008using,reavis2012temporary}. Tinnitus, also known as phantom noise, can occur in people with or without hearing problems. Tinnitus has always been a worldwide health problem, with an overall incidence rate of about 11\% to 16\%~\cite{heller2003classification,nondahl2010ten}. Tinnitus can impair the life quality of a patient in many ways, which may cause sleep problems, difficulty concentrating, and may even develop into psychiatric disorders, such as lethargy, anxiety, and anger~\cite{ nova2006tinitus, espinosa2014tratamiento, eggermont2004neuroscience}. For the treatment and diagnosis of tinnitus, traditional investigators assess with a Visual Analog Scale (VAS) or a temporary questionnaire~\cite{haefeli2006pain,kersten2014pain,powers2015acoustic,adamchic2014reversing}. Haefeli et al.~\cite{haefeli2006pain} and Kersten et al.~\cite{kersten2014pain} designed VAS to subjectively quantify certain sensations of patients using a sliding scale from lack to extreme sensation, e.g., pain. Different to the sliding scale, Powers et al.~\cite{powers2015acoustic} and Adamchic et al.~\cite{adamchic2014reversing} proposed to use questionnaires with only three responses, i.e., yes, sometimes, and no, to identify the tinnitus conditions of sufferers. With the development of Electroencephalogram (EEG), EEG has become a valuable tool to diagnose and treat tinnitus. EEG is capable of recording brain activity and analyzing neural oscillations in real time~\cite{ibarra2020acoustic}. A lot of evidence shows a strong correlation between abnormal EEG and tinnitus. For example, Eggermont et al.~\cite{eggermont2015maladaptive}, and Ibarra-Zarate et al.~\cite{ibarra2020acoustic} suggested that tinnitus may be caused by abnormal restructurings in the tonotopic maps, higher synchrony or spontaneous firing rates in the neuron or the central auditory system. Nguyen et al.~\cite{nguyen2020abnormal} found that tinnitus sufferers have abnormal neural oscillations in the beta and gamma frequency bands.

Traditional EEG-based tinnitus diagnosis and treatment rely on the statistical models to quantify and analyze the provided neurofeedback~\cite{7yearsfeedbackbutwork,focusonsinglebetaband}. For example, Vanneste et al.~\cite{vanneste2015tinnitus} performed a quantitative EEG analysis of 154 patients with tinnitus and used a regression analysis to find that structural differences in gray matter may be associated with the distress and duration of tinnitus. Moazami-Goudarzi et al.~\cite{moazami2010temporo} used power spectra to analyze tinnitus and control groups in a group-wise manner. They found that tinnitus will lead to a power enhancement in the delta and theta band of EEG. Although previous researchers have made significant progress in diagnosing and analyzing tinnitus-related EEG signals, their statistical analysis was case-specific and prone to being trapped in the characteristics of small study populations. To ease such problem, more and more work uses deep learning methods to improve the generalization ability of models~\cite{allgaier2021deep,liu2021generalizable,sun2019multi,zhou2021tinnitus}. For example, Liu et al.~\cite{liu2021generalizable} adopted a Siamese network to augment training samples with contrastive relationships and thus enhance the generalization ability of their model. Allgaier et al.~\cite{allgaier2021deep} used noise reduction and down-sampling to generalize the learning process of their proposed deep learning method and achieved an accuracy of 75.6\% in diagnosing tinnitus. However, these methods are still limited to a single dataset. EEG signals are easy to be influenced by multiple non-physiological factors, e.g., biological or physical conditions of brains~\cite{ibarra2020acoustic}. The data distributions may significantly differ from subjects. Moreover, the different settings of the experimental environment, e.g., time duration and sampling frequency, may lead to deep learning models failing to be applied in a new dataset. Therefore, it is necessary to develop a model that is robust across datasets to handle more diverse and general EEG signals.

\subsection{Cross-dataset EEG Research In Related Fields}
Most current EEG research~\cite{wan2021review,santana2019gp,dai2019domain,zhang2017cross} only takes into account the data distributions across subjects, which still ignores the distribution shift between datasets. Some modern models~\cite{cimtay2020investigating,xu2020cross,rayatdoost2018cross,bird2020cross,lin2019constructing} have evolved to ease the distribution shifts across both subjects and datasets. In other words, these models can be applied in multiple datasets rather than a single dataset. For example, in the field of gesture classification, Bird et al.~\cite{bird2020cross} proposed to transfer knowledge between an EEG dataset recording brainwaves and an Electromyographic (EMG) dataset recording muscular waves based on the pre-trained networks. They achieved a high accuracy of 85.12\% and 93.82\% by transferring EEG-to-EMG and EMG-to-EEG knowledge, respectively. Similarly, for emotion recognition, Cimtay et al.~\cite{cimtay2020investigating} proposed to use a pre-trained neural network from a source dataset to the target dataset and achieved an impressive enhancement in the model performance. These models use pre-trained models to provide extra knowledge that may be meaningful across datasets. To provide a more reliable solution to learn meaningful information across datasets, Xu et al.~\cite{xu2020cross} proposed an online method to pre-align the data formats before training and classification. However, all these methods rely on the annotation of target dataset, while many real-world datasets may lack labels. Therefore, it is necessary to propose a work that can be applied in a more general condition, i.e., only a given source dataset is annotated, and the target dataset that needs classification remains unlabeled.

\subsection{Unsupervised Domain Adaptation}
A domain refers to feature space and the corresponding probability distributions related to a set of data, e.g., a dataset~\cite{pan2009survey}. Unsupervised domain adaptation focuses on the situation where the training and testing data are collected from different distributions, e.g., datasets. The training dataset with labeled data and the testing dataset without any annotation are called the source and target datasets, respectively. Conventional deep learning models may suffer from a significant decrease in the model performance due to the domain difference between source and target datasets~\cite{wilson2020survey}. Unsupervised
domain adaptation aims to ease such performance decrease by adapting the different domains~\cite{lu2015transfer,shao2014transfer,tan2018survey,weiss2016survey}. A lot of works~\cite{chai2016unsupervised,jimenez2020custom,yoo2021transferring,hang2019cross} have been done to adopt unsupervised domain adaptation in the field of EEG. Most of these works only view each subject as a domain and fail to ease the dataset-level domain difference. Limited research~\cite{chai2017fast,lan2018domain} has been done to adopt unsupervised domain adaptation for cross-dataset settings. For example, Lan et al.~\cite{lan2018domain} adopted domain adaptation to reduce both cross-subject and cross-dataset variance based on unsupervised domain adaptation and achieved an impressive improvement in emotion recognition. However, the previous work lacks domain-specific knowledge to handle domain differences related to tinnitus, e.g., the domain difference caused by the different sides of the ear of presentation of the evoking stimulus. Our work is tinnitus-specific domain adaptation that is capable of easing cross-ear and cross-dataset differences.

\subsection{Disentangled Representation}

To better serve specific tasks, many deep-learning-based generative adversarial networks~\cite{kim2018disentangling,tran2017disentangled,lee2018diverse,zhu2020s3vae} disentangle the representations from the given data to learn the specific information. For example, Li et al.~\cite{li2021generalized} used a disentangled auto-encoder to disentangle category-distilling and category-dispersing
factors from images for better category classification. Similar, to EEG signals, there exists more information or noise than tinnitus-related signals in the collected EEG. For example, Cai et al.~\cite{cai2020aberrant} adopted spatial smoothing to increase the signal-to-noise ratio and reduce the influence of signal noise. Joos et al.~\cite{joos2012disentangling} and Meyer et al.~\cite{meyer2014disentangling} used questionnaires to disentangle and study the correlations of depression and distress to EEG from tinnitus patients. Therefore, we propose to disentangle the signals to exclude unrelated factors to obtain the pure classification information for tinnitus diagnosis. 

\begin{figure*}[htb]
\centering
  \includegraphics[width=0.8\textwidth]{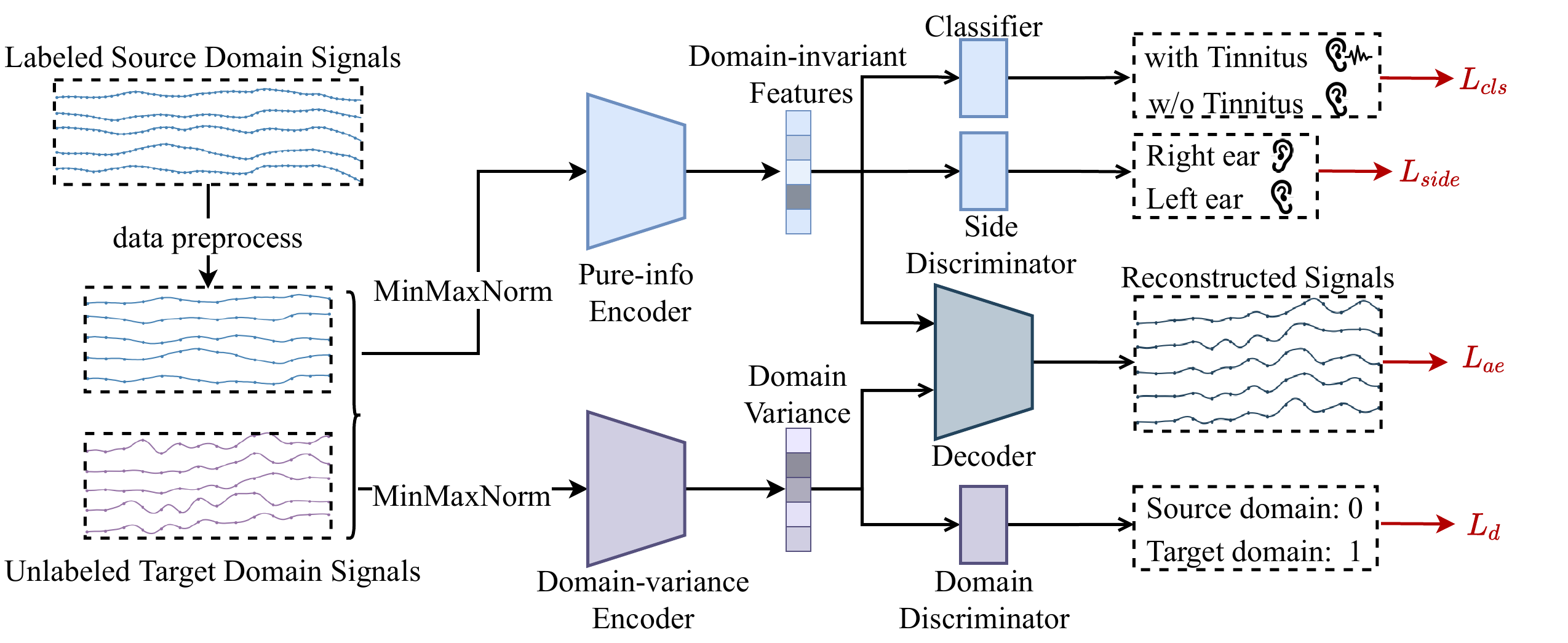}
  \caption{Overview of DSUDA. DSUDA consists of two modules: DAE and DUDA. DAE is composed of Pure-info Encoder, Domain-variance Encoder, Classifier, and Decoder. The labeled data from source domain are first preprocessed to align the unlabeled data from target domain, then the data are normalized to fed into DAE. The two encoders of DAE disentangle the input data into domain-invariant features and domain variance, both of which are then combined as the input of the Decoder. The Classifier takes domain-invariant features as input to predict tinnitus status. DUDA comprises Side Discriminator to distinguish which ear the signal are collected from, and Domain Discriminator to discriminate which domain the data belong to. The two modules are trained together in an adversarial manner to optimize losses $L_{cls}$, $L_{side}$, $L_{ae}$, and $L_d$. In the inference stage, we can directly use the Pure-info Encoder to extract pure class information for prediction.}
  \label{model}
\end{figure*}

\begin{figure}[htb]
\centering
  \includegraphics[width=0.44\textwidth]{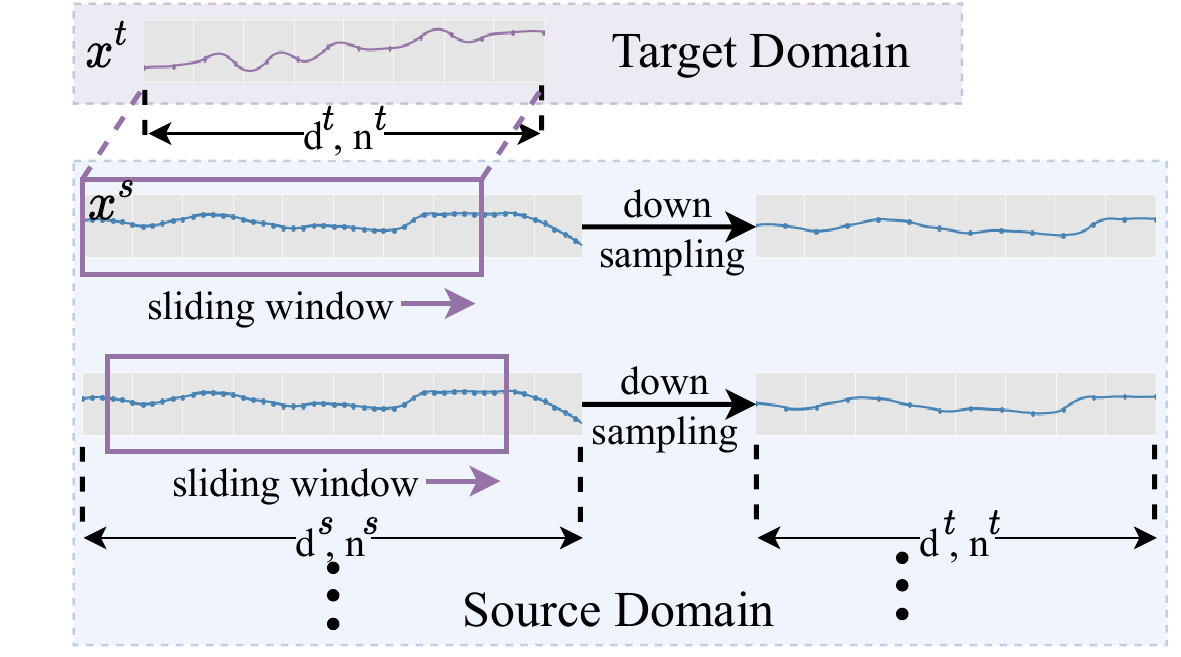}
  \caption{Data preprocessing.}
  \label{data preprocess}
\end{figure}

\section{Method}
\subsection{Problem Definition}

Suppose that dataset $\mathcal{S}=\{(x^{s}, y^{s}, a^{s})|x^{s}\in R^{1\times n^{s}},y^{s}\in\{0,1\},a^{s}\in\{0,1\}\}$ from source domain are given, where $x^{s}=\{o_{j}:j\in[1,n^{s}],o_{j}\in R\}$ is an EEG feature containing signals collected through $n^s$ time points, $y^{s}\in\{0,1\}$ indicates the subject's tinnitus status, and $a^{s}\in\{0,1\}$ implies which ear the signals are collected from. More specifically, $y^{s}=1$ denotes that the subject has tinnitus, and $y^{s}=0$ represents control subject without tinnitus. For $a^{s}$, 0 or 1 denotes left or right ear, respectively. Then, given data from target domain $\mathcal{T}=\{(x^{t}, y^{t}, a^{t})|x^{t}\in R^{1\times n^{t}},y^{t}\in\{0,1\},a^{t}\in\{0,1\}\}$, where $x^{t}$ contains $n^t$ time points, and $y^t$ ($a^t$, respectively) indicates same meanings as $y^s$ ($a^s$, respectively). Our goal is to obtain a general classifier that can diagnose tinnitus for the target domain without labeled target data based on the labeled source data. That is, during training, only $\mathcal{S}$ from the source domain and $\{(x^{t},a^{t})\}$ of the target domain are given, and we need to implement unsupervised domain adaptation to achieve cross-datasets diagnosis.

\subsection{Pipeline Overview}

As Figure~\ref{model} shows, our proposed DSUDA integrates two modules: the Disentangled Auto-Encoder (DAE) and the Side-aware Unsupervised Domain Adaptation (SUDA). DAE (Section~\ref{dae}) disentangles input features from source and target domains into pure class information that are domain-invariant, and domain variance that are diverse noise across datasets caused by different experimental environments. SUDA (Section~\ref{suda}) consists of two-level domain adaptation: the subject-level side adaptation to align features from right and left ears to eliminate the side variance, and the domain-level variance adaptation to exclude domain-variance and noises from pure information for better classification. We accomplish tinnitus diagnosis, feature disentanglement, and domain adaptation based on three kinds of loss functions, namely: (1) reconstruction loss $L_{ae}$, (2) classification loss $L_{cls}$, and (3) domain discrimination loss $L_{side}$ and $L_d$.

\subsection{Data Preprocessing}

Due to the different acquisition equipment and settings of EEG signals, the data in the source and target domains may have significant discrepancies. The overall time duration and time steps of the collection may be different. And the ranges of signal strength or device noise differ across devices. Therefore, it is not feasible to directly apply a classifier suitable for the source domain to the target domain. We need to preprocess the source data to align the data format and scope. 

As shown in Figure~\ref{data preprocess}, suppose that an EEG trial duration from source and target domains, i.e., $x^s$ and $x^t$, last for $d^s$ and $d^t$ milliseconds, respectively. We first adopt the strategy of sliding window to segment $x^s$ into slices $\{x^s_1,x^s_2,...\}$, where the window size equals $d^t$, and the sliding step is a hyper parameter. Then we use down sampling to compress $x^s_i=\{o^s_1,o^s_2,...\}$ to align the number of time points in $x^s_i$ (i.e., $n^w=\frac{d^t*n^s}{d^s}$) and $x^t$ (i.e., $n^t$) as follows:

\begin{gather}\label{slide_down_sample}
\bar{o}^{s}_{p}=\left\{\begin{matrix}
\frac{1}{l}\sum_{q\in [(p-1)*l+1,p*l]}o^{s}_{q} & p\in[1,m]\\ 
\frac{1}{l+1}\sum_{q\in [p*(l+1)-l-m,p*(l+1)-l]}o^{s}_{q} & p\in[m+1,n^{t}]
\end{matrix}\right.\\
s.t. \quad l = \left \lfloor 
\frac{n^{w}}{n^{t}} \right \rfloor \quad m = n^{w}-l*n_{t}
\end{gather}

where $l=\left \lfloor 
\frac{n^{w}}{n^{t}} \right \rfloor$, the floor of $\frac{n^{w}}{n^{t}}$, is the compressing step. In this way, we obtain a set of processed features $\{\bar{x}^s_1,\bar{x}^s_2,...\}$ from the original $x^s$, where $\bar{x}^s_i = \{\bar{o}^{s}_{1},\bar{o}^{s}_{2},...\}$ is consistent with $x^t$ in time duration and time point number. Note that we perform the data preprocessing based on the assumptions that $d^s \geq d^t$ and $n^w \geq n^t$, or we align target domain to source domain in reverse. 
Then, we normalize data from source and target domains with MinMaxNorm separately as follows:

\begin{gather}\label{min_max_norm}
    \hat{o}^{s}_{p}=\frac{\bar{o}^{s}_{p}-\min \bar{o}^{s}_{p}}{\max \bar{o}^{s}_{p} - \min \bar{o}^{s}_{p}}\\
    \hat{o}^{t}_{p}=\frac{o^{t}_{p}-\min o^{t}_{p}}{\max o^{t}_{p} - \min o^{t}_{p}}
\end{gather}
where $o^t_p$ is the $p^{th}$ value in $x^t$. 

\subsection{Disentangled Auto-encoder}\label{dae}

DAE aims to improve tinnitus diagnosis by extracting pure information that excludes categorically irrelevant factors. For this, we build a DAE using two parallel encoders. The purely informative encoder $f_{pe}$ (Pure-info Encoder) learns classification-related features, and the classifier $f_c$ supervises $f_{pe}$ to contain classification information. The domain variance encoder $f_{ve}$ learns class-independent features. The decoder $f_{de}$ constrains $f_{pe}$ to avoid overfitting the classification information, and restricts $f_{pe}$ and $f_{ve}$ from overfitting the source domain.

We take $\hat{\mathcal{S}}=\{(\hat{x}^s, y^s, a^s)\}$ and $\hat{\mathcal{T}}=\{(\hat{x}^t, a^t)\}$ as the input of DAE, where $\hat{x}^s=\{\hat{x^s_1}, \hat{x^s_2},...\}$, $\hat{x^s_1} = \{\hat{o}^{s}_{1},\hat{o}^{s}_{2},...\}$, and $\hat{x}^t = \{\hat{o}^{t}_{1},\hat{o}^{t}_{2},...\}$. Given $\hat{x}\in \mathcal{D}=\mathcal{S}\cup\mathcal{T}$, DAE disentangles $\hat{x}$ into class-related embedding $e^c$ and class-unrelated embedding $e^v$. Then $e^c$ is fed into the classifier $f_c$ for classification, and $e^c$ and $e^v$ are concatenated as input of the decoder $f_{de}$ for reconstruction. DAE is optimized by the classification loss $L_{cls}$ measured by CrossEntropy, and the reconstruction loss $L_{ae}$ calculated by MSE as follows:

\begin{gather}
L_{cls}=CrossEntropy(f_{c}(f_{pe}(\hat{x}^{s})),y^{s}) \label{equ1}
\\
L_{ae}=MSE(f_{de}(f_{pe}(\hat{x}),f_{ve}(\hat{x})),\hat{x}) \label{equ2}
\end{gather}
where $CrossEntropy$ and $MSE$ denote cross-entropy loss and mean square error, respectively.

Note that samples in $\hat{\mathcal{T}}$ are not labeled with tinnitus status. Thus, $L_{ae}$ is calculated based on all $\hat{x}\in \mathcal{D}$, while $L_{cls}$ is computed based on only $\hat{x}^{s} \in \hat{\mathcal{S}}$.

However, only loss $L_{cls}$ and $L_{ae}$ are not sufficient. With $L_{cls}$ and $L_{ae}$, we can only ensure that $f_c$ works for the source domain. To enable its feasibility in the target domain, we hope the extracted class-related information $e^c$ is domain-invariant to generalize to the target domain. Intuitively, after excluding the domain variance from the source domain, the remained information is domain-invariant. Therefore, we can consider the class-unrelated features $e^v$ that should be discarded as the domain variance, and adapt $e^v$ from the source domain to the target domain by the side-aware unsupervised domain adaptation, which will be discussed in Section~\ref{suda}. 

\subsection{Side-aware Unsupervised Domain Adaptation}\label{suda}

The features learned by DAE may not fit target domain. To adapt DAE to the target domain, we propose the SUDA focusing on adapting domain variance from source to target domain. To avoid the confusion caused by different signal patterns of right and left ears, we further equip the SUDA with the side-aware adaptation to align information from both ears.

As shown in Figure~\ref{model}, SUDA is composed of the domain ($f_{dd}$) and side discriminator ($f_{sd}$), predicting which domain the domain-variant features belong to and which ear the domain-invariant features are collected from, respectively. Then, we design the loss functions of the domain and side discriminator as $L_d$ and $L_{side}$, which aims to optimize $f_{dd}$ and $f_{sd}$ to be able to distinguish source domain variance $\hat{x}^s$ as 0, target domain variance $\hat{x}^t$ as 1, left ear features as 0, and right ear features as 1:

\begin{gather}
    L_{d}=\mathbb{E}_{\hat{x}\sim{\mathcal{T}}}[f_{dd}(f_{ve}(\hat{x}))]-\mathbb{E}_{\hat{x}\sim{\mathcal{S}}}[f_{dd}(f_{ve}(\hat{x}))] \label{equ3}
    \\
    L_{side}=\mathbb{E}_{\hat{x}\sim{\mathcal{D}^r}}[f_{sd}(f_{pe}(\hat{x}))]-\mathbb{E}_{\hat{x}\sim{\mathcal{D}^l}}[f_{sd}(f_{pe}(\hat{x}))] \label{equ4}
    \\
    L_{DS} = L_{d}+\eta*L_{side}
\end{gather}
where $\mathcal{D}^r$ and $\mathcal{D}^l$ denote sets of $\hat{x}$ that collected from right and left ears, respectively, $ L_{DS}$ is the overall discriminating loss, and $\eta$ is a hyper-parameter.

With losses $L_d$ and $L_{side}$, we can obtain reliable discriminators. Then, we can optimize DAE by optimizing $f_{ve}$ to confuse the domain discriminator to extract domain variance, and optimizing $f_{pe}$ to confuse the side discriminator to align side information. Intuitively, if the domain discriminators cannot distinguish class-irrelevant features from different domains, the domain variance is adapted successfully and $f_{pe}$ can disentangle class-related information from $\hat{x}$ for both domains. Then, we can apply the classifier trained on source domain to target domain. 

We optimize DAE and SUDA in an adversarial manner as follows: 1) maximizing $L_{DS}$ to enable $f_{dd}$ and $f_{sd}$ to distinguish between different domains and ear sides; 2) optimizing $f_{ve}$, $f_{pe}$, $f_c$, and $f_{de}$ to minimize $L_{DAE}$ to improve the classifier and fool the discriminators. $L_{DAE}$ is the combined loss based on Equations~\ref{equ1}-\ref{equ4} as follows:
\begin{equation}
\begin{split}
   L_{DAE}= &\alpha * L'_{d} + \beta*L'_{side} + L_{ae} +L_{cls}
    \\
    =&-\mathbb{E}_{\hat{x}\sim{\mathcal{T}}}[f_{dd}(f_{ve}(\hat{x}))]+\mathbb{E}_{\hat{x}\sim{\mathcal{D}^l}}[f_{sd}(f_{pe}(\hat{x}))] 
    \\&+ L_{ae} +L_{cls}
\end{split}
\end{equation}
where $\alpha$ and $\beta$ are two  hyper-parameters, $L'_{d}$ and $L'_{side}$ are simplified versions of $L_{d}$ and $L_{side}$ without requirements of recognizing data from source domain or right ear. This is because that optimizing DAE to fit target domain to the source domain and align information from left to right side is sufficient for adaptation between two domains.

\subsection{Training and Inference}

\textbf{Training.} Due to the instability of adversarial training, we first train Pure-info Encoder, Classifier, and Decoder to provide a basic classification ability, and use the trained parameters of Pure-info Encoder as parameters of Domain-variance Encoder. Then, we consider the trained DAE as the initial model, and optimize DAE and SUDA based on $L_{DAE}$ and $L_{DS}$ adversarially.

\textbf{Inference.} We use Pure-info Encoder to extract class-related features from target domain data and use the Classifier to predict tinnitus status directly without modifying its parameters.

\begin{table*}
\centering
\caption{Overall comparisons on both ears with SOTA methods.}
\label{Main_exp_both}
\resizebox{\textwidth}{!}{%
\begin{tabular}{c|ccccccc} 
\toprule
\multirow{2}{*}{Model} & \multicolumn{7}{c}{Both Sides}                                                                          \\ 
\cmidrule{2-8}
                                                                      & NPV          & TNR          & N-F1         & PPV          & TPR          & P-F1         & Acc           \\ 
\midrule
XGBoost                                                               & 0.467(0.000) & 0.525(0.000) & 0.494(0.000) & 0.457(0.000) & 0.400(0.000) & 0.427(0.000) & 0.463(0.000)  \\
Nu-SVC                                                                & 0.467(0.000) & 0.350(0.000) & 0.400(0.000) & 0.480(0.000) & 0.600(0.000) & 0.533(0.000) & 0.475(0.000)  \\
nCSP                                                                  & 0.568(0.055) & 0.625(0.078) & 0.595(0.058) & 0.583(0.054) & 0.525(0.085) & 0.553(0.063) & 0.575(0.053)  \\
DeepNet                                                               & 0.714(0.243) & 0.125(0.042) & 0.213(0.070) & 0.521(0.010) & 0.950(0.019) & 0.673(0.009) & 0.538(0.018)  \\
ShalowNet                                                             & 0.581(0.044) & 0.625(0.147) & 0.602(0.112) & 0.595(0.033) & 0.550(0.110) & 0.571(0.040) & 0.588(0.035)  \\
AE-XGB                                                                & 0.696(0.071) & 0.400(0.170) & 0.508(0.158) & 0.579(0.033) & 0.825(0.126) & 0.680(0.041) & 0.613(0.041)  \\
EEGNET                                                                & 0.605(0.081) & 0.650(0.230) & 0.627(0.181) & 0.622(0.048) & 0.575(0.159) & 0.597(0.055) & 0.613(0.049)  \\
AE                                                                    & 0.660(0.047) & 0.775(0.077) & 0.713(0.032) & 0.727(0.104) & 0.600(0.154) & 0.658(0.143) & 0.688(0.056)  \\
SAE                                                                   & 0.703(0.029) & 0.650(0.063) & 0.675(0.018) & 0.674(0.027) & 0.725(0.076) & 0.699(0.034) & 0.688(0.013)  \\
\midrule

UDA   & 0.833(0.041) & 0.500(0.118) & 0.625(0.050) & 0.643(0.050) & 0.900(0.105) & 0.750(0.035) & 0.700(0.018)  \\
SUDA  & 0.744(0.022) & 0.725(0.051) & 0.734(0.017) & 0.732(0.030) & 0.750(0.054) & 0.741(0.022) & 0.738(0.012)  \\
DSUDA & 0.789(0.031) & 0.750(0.057) & 0.769(0.024) & 0.762(0.032) & 0.800(0.057) & 0.780(0.023) & 0.775(0.018)  \\
\bottomrule
\end{tabular}
}
\end{table*}

\begin{table*}
\centering
\caption{Results on the left ear.}
\label{Main_exp_left}
\resizebox{\textwidth}{!}{%
\begin{tabular}{c|ccccccc} 
\toprule
\multirow{2}{*}{\begin{tabular}[c]{@{}c@{}}\\Model\end{tabular}} & \multicolumn{7}{c}{Left Side}                                                                          \\ 
\cmidrule{2-8}
                                                                 & NPV          & TNR          & N-F1         & PPV          & TPR          & P-F1         & Acc           \\ 
\midrule
XGBoost                                                          & 0.455(0.000) & 0.500(0.000) & 0.476(0.000) & 0.444(0.000) & 0.400(0.000) & 0.421(0.000) & 0.450(0.000)  \\
Nu-SVC                                                           & 0.467(0.000) & 0.350(0.000) & 0.400(0.000) & 0.480(0.000) & 0.600(0.000) & 0.533(0.000) & 0.475(0.000)  \\
nCSP                                                             & 0.591(0.088) & 0.650(0.109) & 0.619(0.090) & 0.611(0.078) & 0.550(0.114) & 0.579(0.091) & 0.600(0.081)  \\
DeepNet                                                          & 0.500(0.245) & 0.100(0.053) & 0.167(0.085) & 0.500(0.016) & 0.900(0.026) & 0.643(0.017) & 0.575(0.029)  \\
ShalowNet                                                        & 0.524(0.063) & 0.550(0.136) & 0.537(0.110) & 0.526(0.036) & 0.500(0.122) & 0.513(0.055) & 0.525(0.044)  \\
AE-XGB                                                           & 0.615(0.143) & 0.400(0.179) & 0.485(0.169) & 0.556(0.049) & 0.750(0.146) & 0.638(0.065) & 0.575(0.060)  \\
EEGNET                                                           & 0.600(0.115) & 0.600(0.240) & 0.600(0.196) & 0.600(0.054) & 0.600(0.169) & 0.600(0.068) & 0.600(0.062)  \\
AE                                                               & 0.652(0.053) & 0.750(0.078) & 0.698(0.043) & 0.706(0.157) & 0.600(0.159) & 0.649(0.161) & 0.675(0.069)  \\
SAE                                                              & 0.750(0.038) & 0.600(0.077) & 0.667(0.035) & 0.667(0.036) & 0.800(0.073) & 0.727(0.030) & 0.700(0.023)  \\ 
\midrule
UDA   & 0.900(0.063) & 0.450(0.137) & 0.600(0.075) & 0.633(0.072) & 0.950(0.111) & 0.760(0.051) & 0.700(0.047)  \\
SUDA  & 0.737(0.028) & 0.700(0.054) & 0.718(0.027) & 0.714(0.034) & 0.750(0.054) & 0.732(0.028) & 0.725(0.023)  \\
DSUDA & 0.833(0.051) & 0.750(0.070) & 0.789(0.045) & 0.773(0.042) & 0.850(0.064) & 0.810(0.037) & 0.800(0.037)  \\
\bottomrule
\end{tabular}
}
\end{table*}

\begin{table*}
\centering
\caption{Results on the right ear.}
\label{Main_exp_right}
\resizebox{\textwidth}{!}{%
\begin{tabular}{c|ccccccc} 
\toprule
\multirow{2}{*}{\begin{tabular}[c]{@{}c@{}}\\Model\end{tabular}} & \multicolumn{7}{c}{Right Side}                                                                          \\ 
\cmidrule{2-8}
                                                                 & NPV          & TNR          & N-F1         & PPV          & TPR          & P-F1         & Acc           \\ 
\midrule
XGBoost                                                          & 0.478(0.000) & 0.550(0.000) & 0.512(0.000) & 0.471(0.000) & 0.400(0.000) & 0.432(0.000) & 0.475(0.000)  \\
Nu-SVC                                                           & 0.467(0.000) & 0.350(0.000) & 0.400(0.000) & 0.480(0.000) & 0.600(0.000) & 0.533(0.000) & 0.475(0.000)  \\
nCSP                                                             & 0.545(0.068) & 0.600(0.099) & 0.571(0.077) & 0.556(0.062) & 0.500(0.109) & 0.526(0.078) & 0.550(0.063)  \\
DeepNet                                                          & 1.000(0.416) & 0.150(0.047) & 0.261(0.082) & 0.541(0.011) & 1.000(0.028) & 0.702(0.012) & 0.500(0.021)  \\
ShalowNet                                                        & 0.636(0.049) & 0.700(0.166) & 0.667(0.125) & 0.667(0.045) & 0.600(0.104) & 0.632(0.033) & 0.650(0.043)  \\
AE-XGB                                                           & 0.800(0.115) & 0.400(0.174) & 0.533(0.164) & 0.600(0.039) & 0.900(0.127) & 0.720(0.044) & 0.650(0.051)  \\
EEGNET                                                           & 0.609(0.060) & 0.700(0.226) & 0.651(0.170) & 0.647(0.054) & 0.550(0.155) & 0.595(0.050) & 0.625(0.046)  \\
AE                                                               & 0.667(0.049) & 0.800(0.098) & 0.727(0.037) & 0.750(0.110) & 0.600(0.160) & 0.667(0.140) & 0.700(0.055)  \\
SAE                                                              & 0.667(0.038) & 0.700(0.063) & 0.683(0.027) & 0.684(0.037) & 0.650(0.090) & 0.667(0.054) & 0.675(0.031)  \\ 
\midrule
UDA   & 0.786(0.064) & 0.550(0.120) & 0.647(0.056) & 0.654(0.063) & 0.850(0.126) & 0.739(0.065) & 0.700(0.043)  \\
SUDA  & 0.750(0.031) & 0.750(0.060) & 0.750(0.023) & 0.750(0.042) & 0.750(0.070) & 0.750(0.036) & 0.750(0.023)  \\
DSUDA & 0.750(0.046) & 0.750(0.074) & 0.750(0.037) & 0.750(0.056) & 0.750(0.082) & 0.750(0.047) & 0.750(0.037)  \\
\bottomrule
\end{tabular}
}
\end{table*}

\begin{figure*}[htb]
\centering % <-- added
\begin{subfigure}{0.32\textwidth}
  \includegraphics[width=\textwidth]{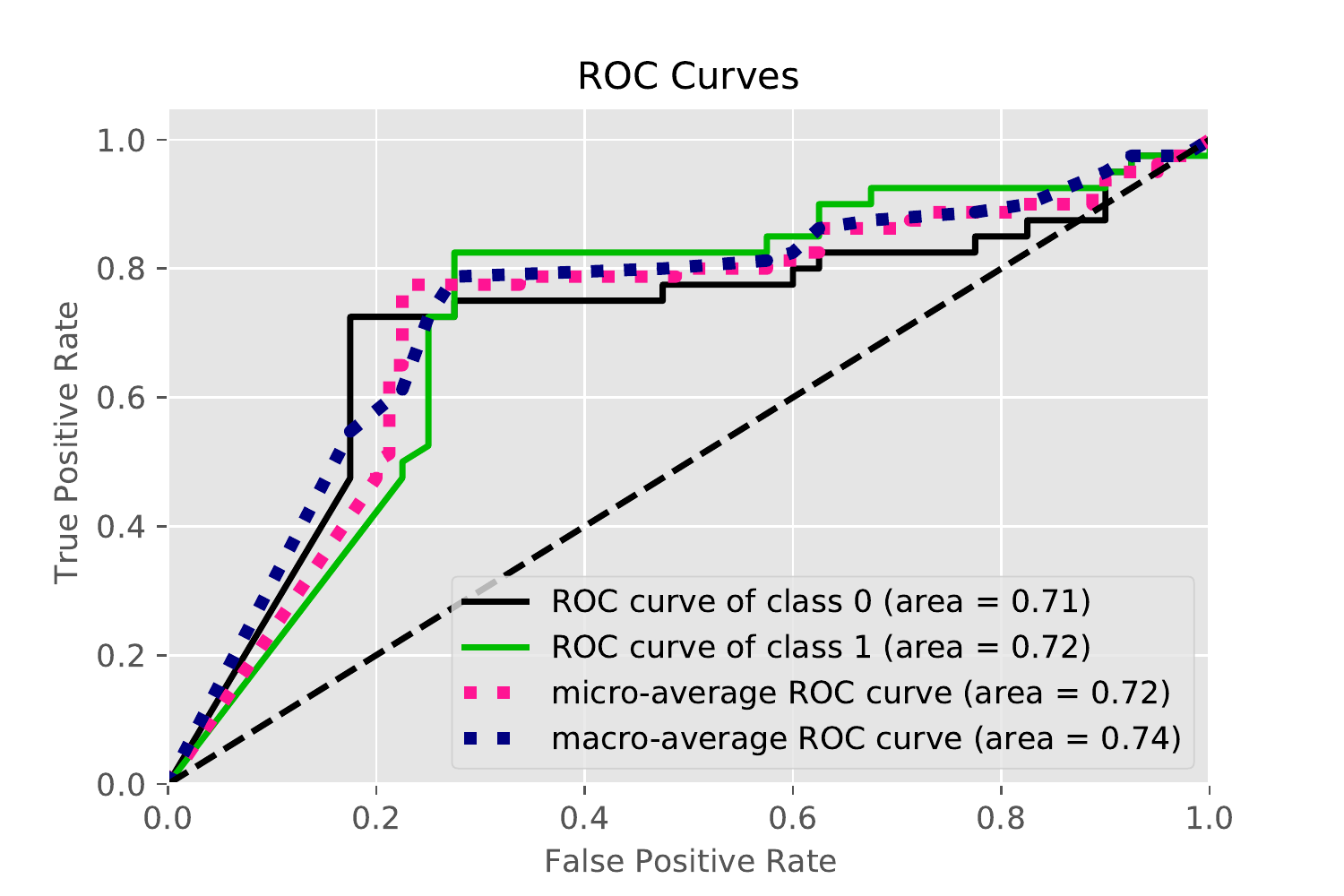}
    \centering
  \caption{ROC curve of both sides.}
\end{subfigure}\hfil % <-- added
\begin{subfigure}{0.32\textwidth}
  \includegraphics[width=\textwidth]{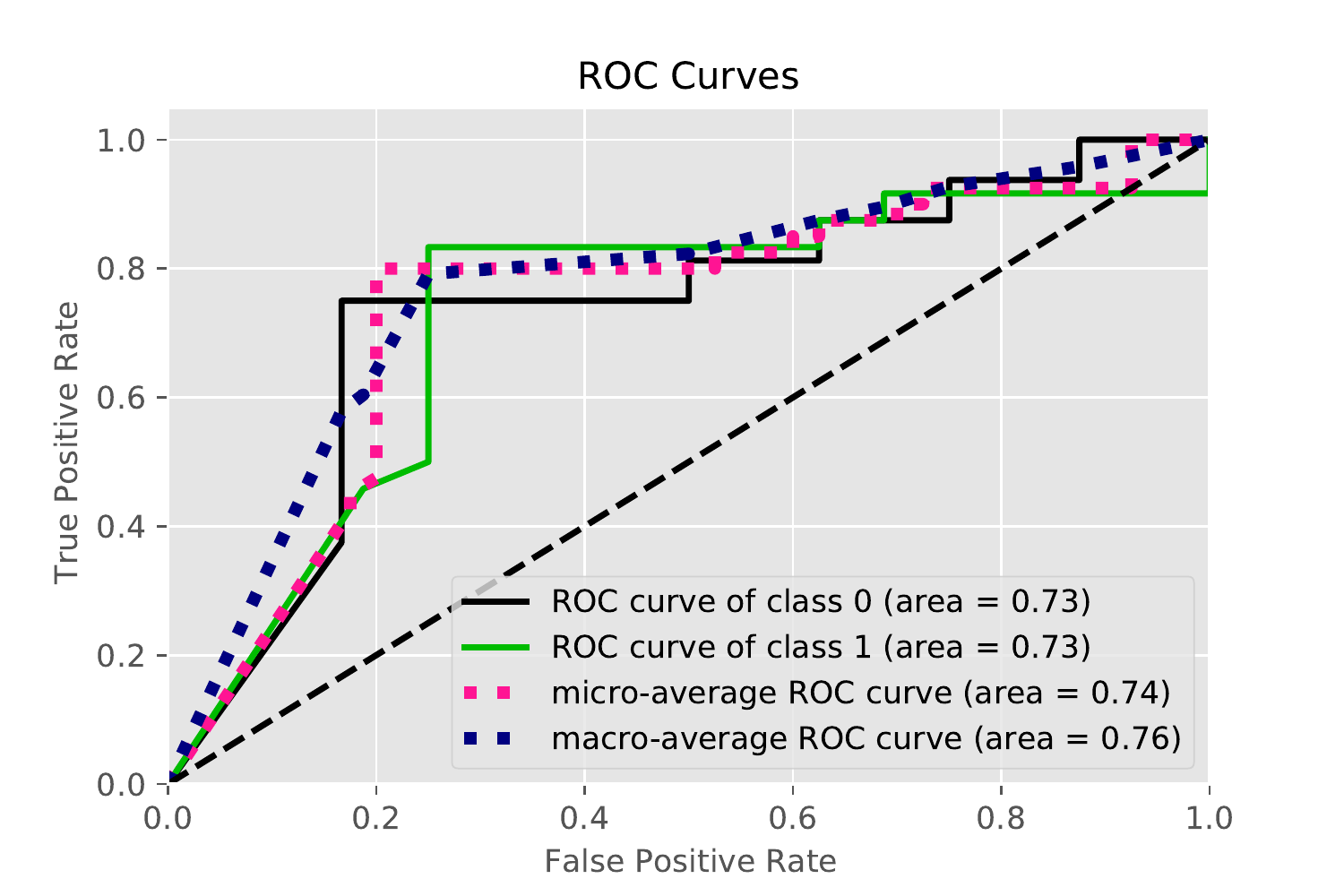}
    \centering
  \caption{ROC curve of left side.}
\end{subfigure}\hfil % <-- added
\begin{subfigure}{0.32\textwidth}
  \includegraphics[width=\textwidth]{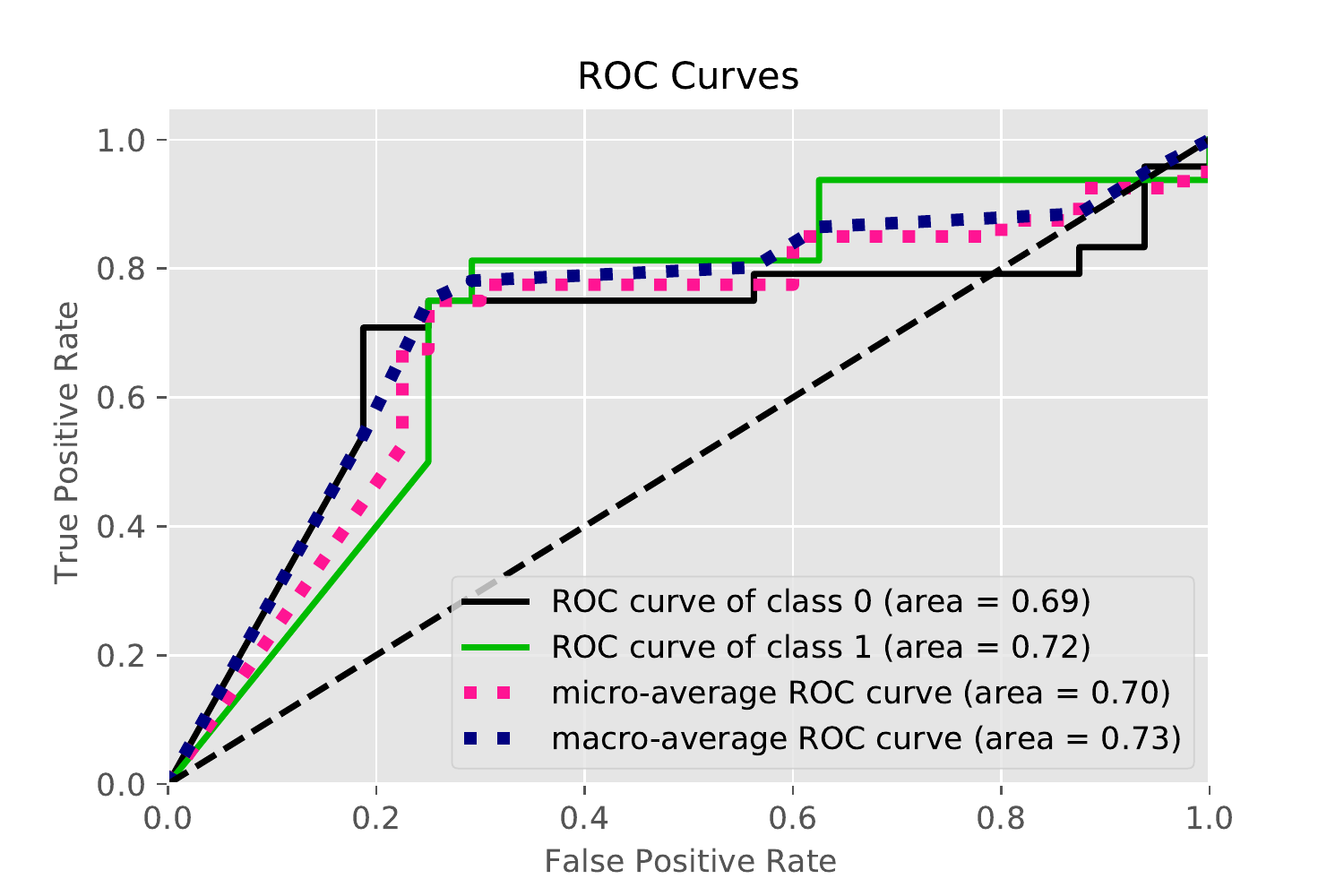}
    \centering
  \caption{ROC curve of right side.}
\end{subfigure}\hfil % <-- added
\caption{ROC curves of DSUDA. Class 0 and 1 represent subjects with or without subjective tinnitus, respectively.}\label{ROC_curve}
\end{figure*}

\begin{figure*}[htb]
\centering % <-- added
\begin{subfigure}{0.23\textwidth}
  \includegraphics[width=\textwidth]{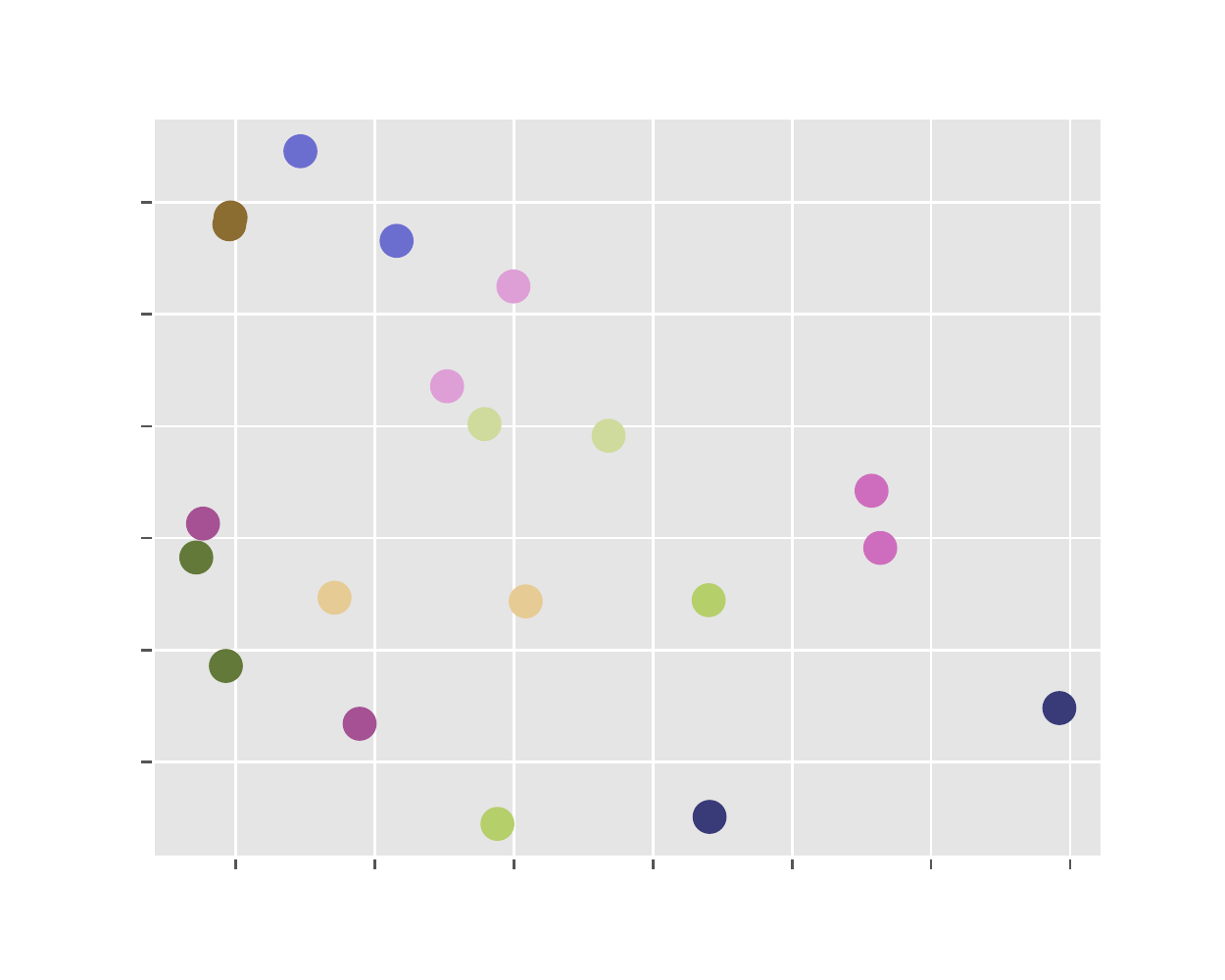}
    \centering
  \caption{Raw signals.}
\end{subfigure}\hfil % <-- added
\begin{subfigure}{0.23\textwidth}
  \includegraphics[width=\textwidth]{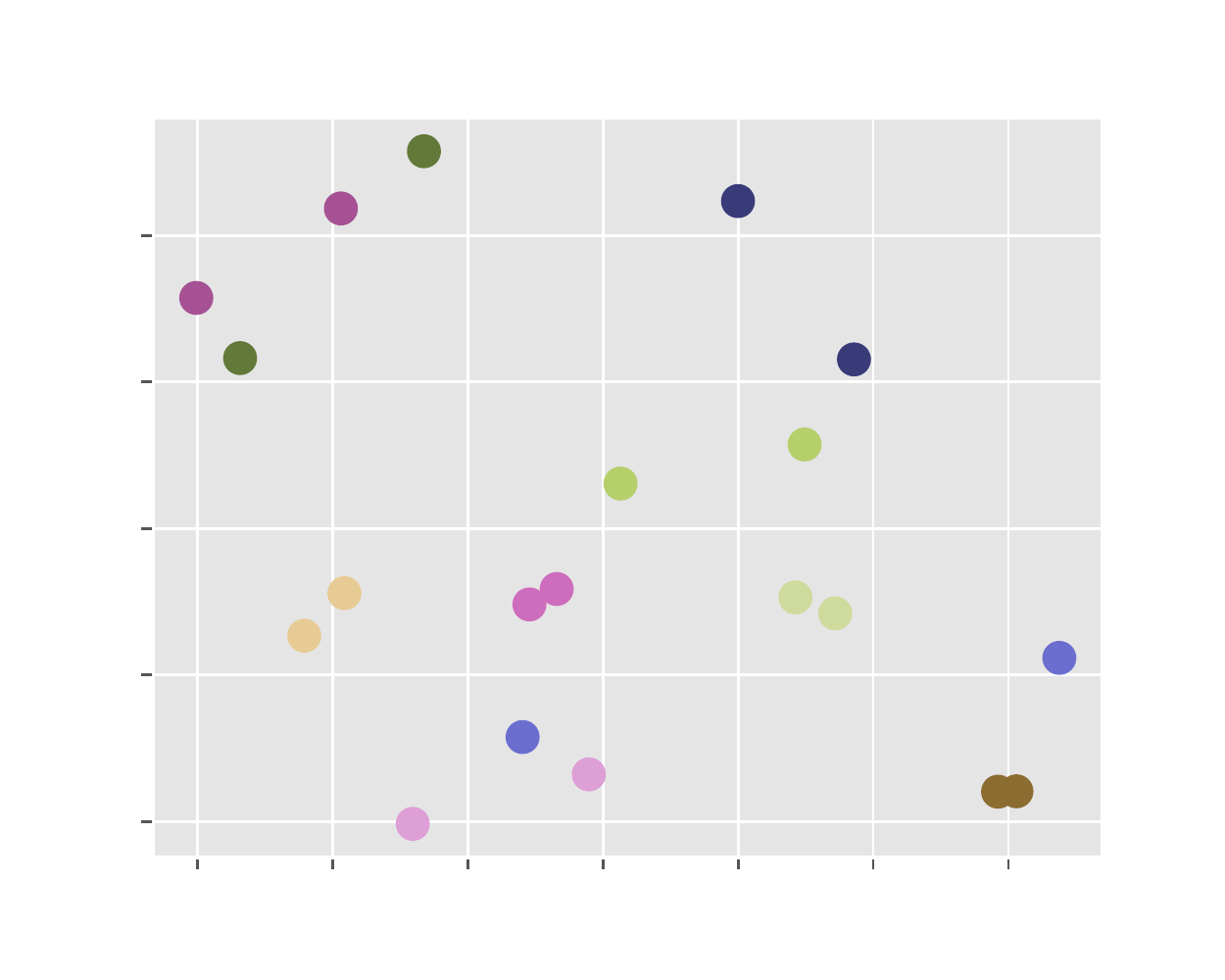}
    \centering
  \caption{Features learned by SUDA.}
\end{subfigure}\hfil % <-- added
\begin{subfigure}{0.23\textwidth}
  \includegraphics[width=\textwidth]{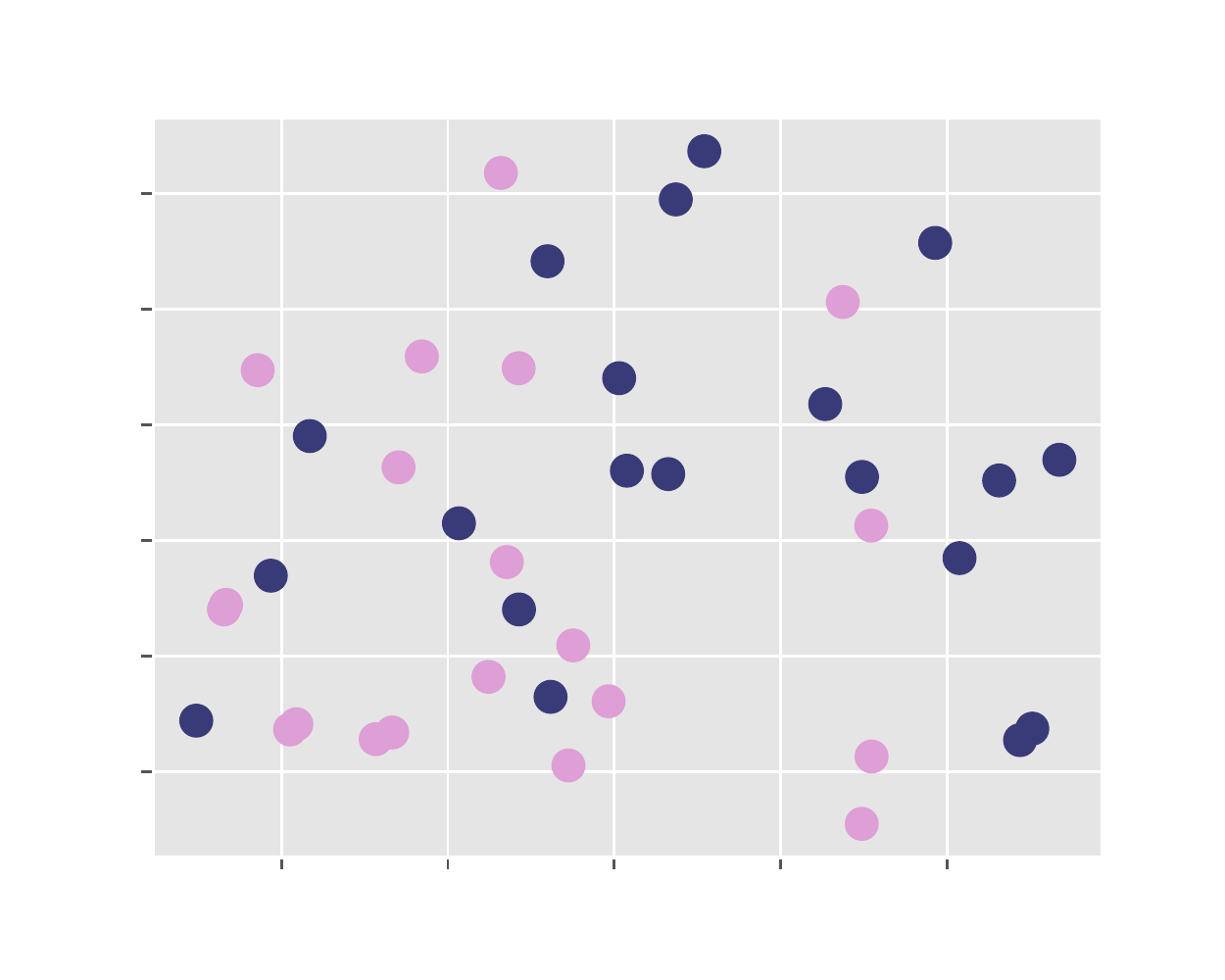}
    \centering
  \caption{Raw signals.}
\end{subfigure}\hfil % <-- added
\begin{subfigure}{0.23\textwidth}
  \includegraphics[width=\textwidth]{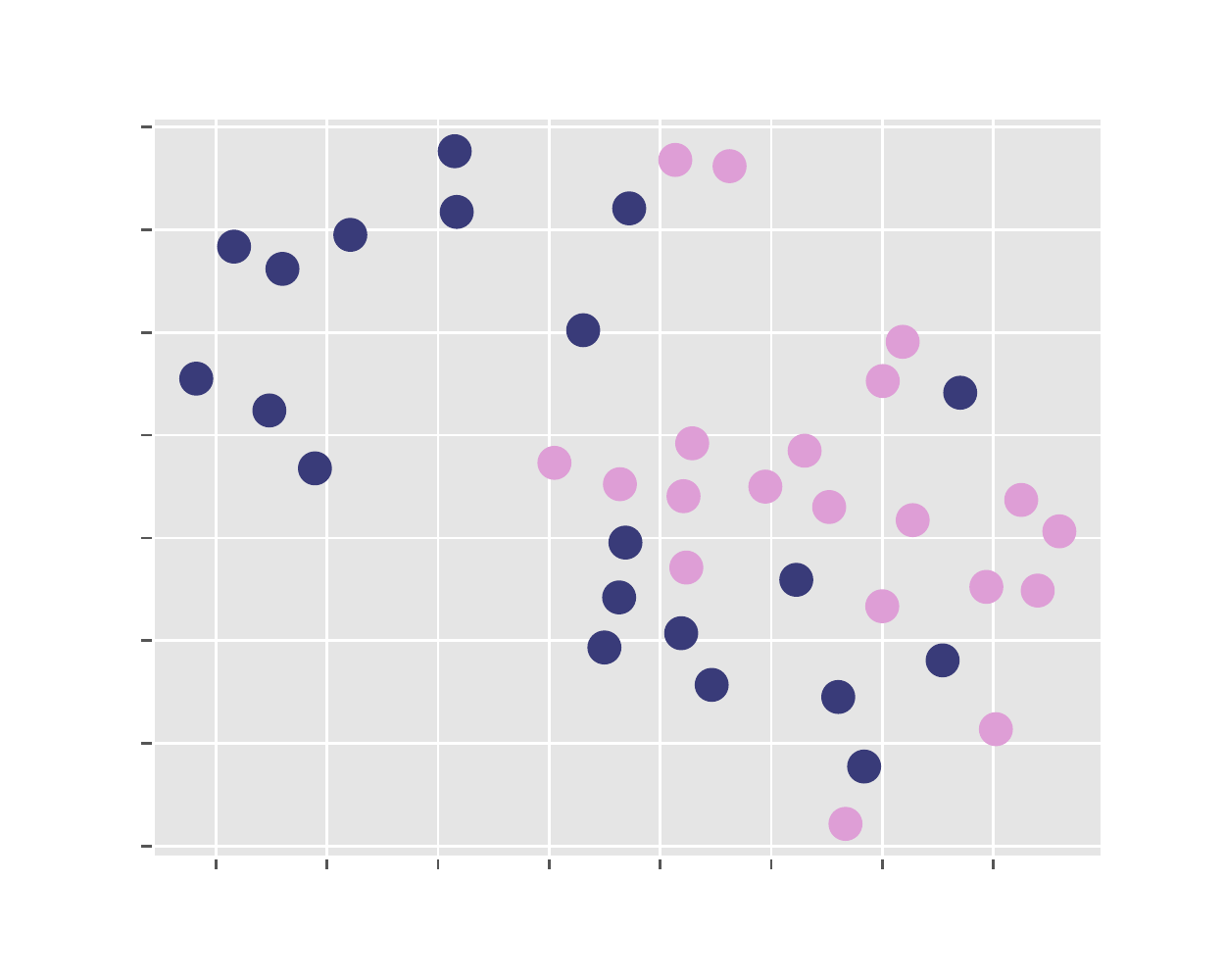}
    \centering
  \caption{Features learned by $f_{pe}$ in DSUDA.}
\end{subfigure}\hfil % <-- added
\caption{Visualization of SUDA and DSUDA. (a)-(b): Raw signals and features learned by SUDA of 10 subjects from target domain. Every subject has two signals, each for one ear. Signals belonging to the same subjected are marked in the same color. (c)-(d) Raw signals and disentangled domain-invariant features learned by DSUDA in target domain of 20 subjects. Tinnitus samples are in pink; control samples are in blue.}
\label{embedding}
\end{figure*}

\section{Experiments}
\subsection{Dataset and Experimental Setting}
We evaluate our model based on two Auditory brainstem responses (ABRs) datasets~\cite{schaette2011tinnitus,guest2017tinnitus} based on EEG sensors and consider the larger datasets~\cite{schaette2011tinnitus} as the source domain and the smaller one~\cite{guest2017tinnitus} as the target domain. ABRs are signals that originate from the early stages of the auditory pathway and evoked potentials recorded with EEG sensors~\cite{schaette2011tinnitus,guest2017tinnitus}. Schaette and McAlpine~\cite{schaette2011tinnitus} used presentation levels of 90 and 100 dB peSPL with an 11 click/s rate for the source domain approved by the University College London (UCL) ethics committee. The ethics ID number is 2039/002 and the declaration is the Helsinki declaration. Guest et al.~\cite{guest2017tinnitus} collected the target domain in response to 102 dB peSPL clicks presented with a rate of 7 clicks/s approved by the National Research Ethics Service Greater Manchester West Ethics Committee with REC reference 15/NW/0133 and IRAS project ID 168221. The signals are sampled with 50 kHz and bandpass-filtered before 100 Hz and 1500 Hz. The source dataset contains 408 EEG signals of 38 subjects recorded while an acoustic click train was presented over headphones to the affected ears for patients or either ear for control subjects, i.e., only one ear of each subject is recorded. Each trial lasts for 10 milliseconds resulting in 500-time points. The target dataset collects 80 EEG signals of 40 subjects with both of their ears recorded. The time duration is 8 milliseconds with 131-time points per trial.

This target dataset was approved by the University College London (UCL) ethics committee. The ethics ID number of our study is 2039/002. The declaration is the Helsinki declaration.

The 2017 study was approved by the National Research Ethics Service Greater Manchester West Ethics Committee, REC reference 15/NW/0133, IRAS project ID 168221.

We use Fully Connected (FC) Layers to build our model. $f_{pe}$ and $f_{ve}$ have the same structures consisting of 2 FC layers with Tanh as the activation function. Other model components also contain 2 FC layers while adopting Sigmoid as the activation function. During data preprocessing, we set the sliding step as 20; thus an original trial from the source domain can be sliced into 6 segments, which are further down-sampled. We set learning rates of DAE and SUDA as 1.0e-3 with different updating steps per batch, 10 and 1, respectively. We set the loss ratios $\eta$, $\alpha$, and $\beta$ as 1.0, 1.0, and 1.0e-6.

We choose 9 machine or deep learning methods as competitors: XGBoost~\cite{chen2016xgboost},  nu-SVC~\cite{v-svm},  nCSP~\cite{ncsp},  DeepNet \cite{DeepConvNet}, ShallowNet \cite{DeepConvNet}, AEXGB \cite{AEXGB}, EEGNet \cite{eegnet},  AE~\cite{sunilkumar2021bio}, and SAE~\cite{liu2021generalizable}. Unlike regular classification tasks, incorrect tinnitus diagnosis may cause severe consequences. Accuracy alone is not sufficient to evaluate the models. Therefore, we adopt Accuracy(Acc) along with Negative Predictive Value (NPV), True Negative Rate (TNR), Positive Predictive Value (PPV), True Positive Rate (TPR), F1-score for negative samples (N-F1), and F1-score for positive samples (P-F1)  to evaluate the performance comprehensively. 

\subsection{Experimental Results}
\subsubsection{Comparison with State-of-the-art Methods}
We present the results of our model and comparison in 3 different settings: both-side, left-side, and right-side, respectively, in Tables~(\ref{Main_exp_both}-\ref{Main_exp_right}). Among the compared methods, generative-based models, e.g., SAE, perform better than non-generative models, suggesting that the generative algorithms have a better generalization ability. DSUDA performs the best among all the algorithms, outperforming the second-best algorithm (including our variant) on N-F1, P-F1, accuracy by 3.5\%/7.1\%, 3.0\%/5.0\%, and 3.7\%/7.5\% in both-/left-side settings, respectively; 2.3\%, 3.0\%, and 5.0\% higher than the best comparison algorithm on the right-side setting, respectively. Our model shows better improvement in the left ear than in other settings, which indicates that the left ear presentations have more transferable knowledge on these two datasets than the right ear. UDA, SUDA, and DSUDA can improve performance on both left-side and both-side settings, indicating that domain adaptation, side-aware domain adaptation, and cross-dataset domain adaptation are effective in enhancing the model performance. On the right ear setting, the improvement on DSUDA is limited, which shows that the side-aware domain adaptation is more generalized and can fit more situations than the dataset domain adaptation.

\begin{table*}[h]
\centering
\caption{Parameter study of DSUDA on different updating steps of DAE.}
    \label{Ablation_k2}
    \resizebox{\textwidth}{!}{%
\begin{tabular}{c|ccccc|ccccc|ccccc}
\toprule
\multirow{2}{*}{Updating Steps} & \multicolumn{5}{c|}{Both Sides}        & \multicolumn{5}{c|}{Left Side}         & \multicolumn{5}{c}{Right Side}         \\\cmidrule{2-16}
                            & NPV   & TNR   & PPV   & TPR   & Acc   & NPV   & TNR   & PPV   & TPR   & Acc   & NPV   & TNR   & PPV   & TPR   & Acc    \\
\midrule
1  & 0.725 & 0.725 & 0.725 & 0.725 & 0.725 & 0.737 & 0.700 & 0.714 & 0.750 & 0.725 & 0.714 & 0.750 & 0.737 & 0.700 & 0.725  \\
2  & 0.730 & 0.675 & 0.698 & 0.750 & 0.713 & 0.750 & 0.600 & 0.667 & 0.800 & 0.700 & 0.714 & 0.750 & 0.737 & 0.700 & 0.725  \\
3  & 0.681 & 0.800 & 0.758 & 0.625 & 0.713 & 0.714 & 0.750 & 0.737 & 0.700 & 0.725 & 0.654 & 0.850 & 0.786 & 0.550 & 0.700  \\
4  & 0.700 & 0.700 & 0.700 & 0.700 & 0.700 & 0.684 & 0.650 & 0.667 & 0.700 & 0.675 & 0.714 & 0.750 & 0.737 & 0.700 & 0.725  \\
5  & 0.705 & 0.775 & 0.750 & 0.675 & 0.725 & 0.684 & 0.650 & 0.667 & 0.700 & 0.675 & 0.720 & 0.900 & 0.867 & 0.650 & 0.775  \\
6  & 0.750 & 0.675 & 0.705 & 0.775 & 0.725 & 0.800 & 0.600 & 0.680 & 0.850 & 0.725 & 0.714 & 0.750 & 0.737 & 0.700 & 0.725  \\
7  & 0.707 & 0.725 & 0.718 & 0.700 & 0.713 & 0.737 & 0.700 & 0.714 & 0.750 & 0.725 & 0.682 & 0.750 & 0.722 & 0.650 & 0.700  \\
8  & 0.707 & 0.725 & 0.718 & 0.700 & 0.713 & 0.700 & 0.700 & 0.700 & 0.700 & 0.700 & 0.714 & 0.750 & 0.737 & 0.700 & 0.725  \\
9  & 0.732 & 0.750 & 0.744 & 0.725 & 0.738 & 0.737 & 0.700 & 0.714 & 0.750 & 0.725 & 0.727 & 0.800 & 0.778 & 0.700 & 0.750  \\
10 & 0.789 & 0.750 & 0.762 & 0.800 & 0.775 & 0.833 & 0.750 & 0.773 & 0.850 & 0.800 & 0.750 & 0.750 & 0.750 & 0.750 & 0.750  \\
\bottomrule
\end{tabular}
}
\end{table*}

\begin{table*}[h]
\centering
\caption{Parameter study of DSUDA on different updating steps of SUDA.}
    \label{Ablation_k1}
    \resizebox{\textwidth}{!}{%
\begin{tabular}{c|ccccc|ccccc|ccccc}
\toprule
\multirow{2}{*}{Updating Steps} & \multicolumn{5}{c|}{Both Sides}        & \multicolumn{5}{c|}{Left Side}         & \multicolumn{5}{c}{Right Side}         \\\cmidrule{2-16}
                            & NPV   & TNR   & PPV   & TPR   & Acc   & NPV   & TNR   & PPV   & TPR   & Acc   & NPV   & TNR   & PPV   & TPR   & Acc    \\
\midrule
1  & 0.725 & 0.725 & 0.725 & 0.725 & 0.725 & 0.737 & 0.700 & 0.714 & 0.750 & 0.725 & 0.714 & 0.750 & 0.737 & 0.700 & 0.725  \\
2  & 0.673 & 0.825 & 0.774 & 0.600 & 0.713 & 0.667 & 0.800 & 0.750 & 0.600 & 0.700 & 0.680 & 0.850 & 0.800 & 0.600 & 0.725  \\
3  & 0.718 & 0.700 & 0.707 & 0.725 & 0.713 & 0.684 & 0.650 & 0.667 & 0.700 & 0.675 & 0.750 & 0.750 & 0.750 & 0.750 & 0.750  \\
4  & 0.718 & 0.700 & 0.707 & 0.725 & 0.713 & 0.684 & 0.650 & 0.667 & 0.700 & 0.675 & 0.750 & 0.750 & 0.750 & 0.750 & 0.750  \\
5  & 0.737 & 0.700 & 0.714 & 0.750 & 0.725 & 0.722 & 0.650 & 0.682 & 0.750 & 0.700 & 0.750 & 0.750 & 0.750 & 0.750 & 0.750  \\
6  & 0.743 & 0.650 & 0.689 & 0.775 & 0.713 & 0.765 & 0.650 & 0.696 & 0.800 & 0.725 & 0.722 & 0.650 & 0.682 & 0.750 & 0.700  \\
7  & 0.725 & 0.725 & 0.725 & 0.725 & 0.725 & 0.737 & 0.700 & 0.714 & 0.750 & 0.725 & 0.714 & 0.750 & 0.737 & 0.700 & 0.725  \\
8  & 0.725 & 0.725 & 0.725 & 0.725 & 0.725 & 0.737 & 0.700 & 0.714 & 0.750 & 0.725 & 0.714 & 0.750 & 0.737 & 0.700 & 0.725  \\
9  & 0.725 & 0.725 & 0.725 & 0.725 & 0.725 & 0.700 & 0.700 & 0.700 & 0.700 & 0.700 & 0.750 & 0.750 & 0.750 & 0.750 & 0.750  \\
10 & 0.711 & 0.800 & 0.771 & 0.675 & 0.738 & 0.737 & 0.700 & 0.714 & 0.750 & 0.725 & 0.692 & 0.900 & 0.857 & 0.600 & 0.750  \\
\bottomrule
\end{tabular}
}
\end{table*}

\begin{table*}[h]
\centering
\caption{Parameter study of DSUDA on different learning rates of DAE.}
    \label{Ablation_lr2}
    \resizebox{\textwidth}{!}{%
\begin{tabular}{c|ccccc|ccccc|ccccc}
\toprule
\multirow{2}{*}{learning rate} & \multicolumn{5}{c|}{Both Sides}        & \multicolumn{5}{c|}{Left Side}         & \multicolumn{5}{c}{Right Side}         \\\cmidrule{2-16}
                            & NPV   & TNR   & PPV   & TPR   & Acc   & NPV   & TNR   & PPV   & TPR   & Acc   & NPV   & TNR   & PPV   & TPR   & Acc    \\
\midrule
6.0e-4 & 0.690 & 0.725 & 0.711 & 0.675 & 0.700 & 0.667 & 0.700 & 0.684 & 0.650 & 0.675 & 0.714 & 0.750 & 0.737 & 0.700 & 0.725  \\
7.0e-4 & 0.682 & 0.750 & 0.722 & 0.650 & 0.700 & 0.714 & 0.750 & 0.737 & 0.700 & 0.725 & 0.652 & 0.750 & 0.706 & 0.600 & 0.675  \\
8.0e-4 & 0.714 & 0.750 & 0.737 & 0.700 & 0.725 & 0.750 & 0.750 & 0.750 & 0.750 & 0.750 & 0.682 & 0.750 & 0.722 & 0.650 & 0.700  \\
9.0e-4 & 0.714 & 0.750 & 0.737 & 0.700 & 0.725 & 0.750 & 0.750 & 0.750 & 0.750 & 0.750 & 0.682 & 0.750 & 0.722 & 0.650 & 0.700  \\
1.0e-3 & 0.689 & 0.775 & 0.743 & 0.650 & 0.713 & 0.667 & 0.700 & 0.684 & 0.650 & 0.675 & 0.708 & 0.850 & 0.813 & 0.650 & 0.750  \\
1.1e-3 & 0.714 & 0.750 & 0.737 & 0.700 & 0.725 & 0.750 & 0.750 & 0.750 & 0.750 & 0.750 & 0.682 & 0.750 & 0.722 & 0.650 & 0.700  \\
1.2e-3 & 0.714 & 0.750 & 0.737 & 0.700 & 0.725 & 0.750 & 0.750 & 0.750 & 0.750 & 0.750 & 0.682 & 0.750 & 0.722 & 0.650 & 0.700  \\
1.3e-3 & 0.667 & 0.800 & 0.750 & 0.600 & 0.700 & 0.682 & 0.750 & 0.722 & 0.650 & 0.700 & 0.654 & 0.850 & 0.786 & 0.550 & 0.700  \\
1.4e-3 & 0.757 & 0.700 & 0.721 & 0.775 & 0.738 & 0.824 & 0.700 & 0.739 & 0.850 & 0.775 & 0.700 & 0.700 & 0.700 & 0.700 & 0.700  \\
1.5e-3 & 0.757 & 0.700 & 0.721 & 0.775 & 0.738 & 0.778 & 0.700 & 0.727 & 0.800 & 0.750 & 0.737 & 0.700 & 0.714 & 0.750 & 0.725  \\
\bottomrule
\end{tabular}
}
\end{table*}

\begin{table*}[h]
\centering
\caption{Parameter study of DSUDA on different learning rates of SUDA.}
    \label{Ablation_lr1}
    \resizebox{\textwidth}{!}{%
\begin{tabular}{c|ccccc|ccccc|ccccc}
\toprule
\multirow{2}{*}{learning rate} & \multicolumn{5}{c|}{Both Sides}        & \multicolumn{5}{c|}{Left Side}         & \multicolumn{5}{c}{Right Side}         \\\cmidrule{2-16}
                            & NPV   & TNR   & PPV   & TPR   & Acc   & NPV   & TNR   & PPV   & TPR   & Acc   & NPV   & TNR   & PPV   & TPR   & Acc    \\
\midrule
6.0e-4 & 0.680 & 0.850 & 0.800 & 0.600 & 0.725 & 0.708 & 0.850 & 0.813 & 0.650 & 0.750 & 0.654 & 0.850 & 0.786 & 0.550 & 0.700  \\
7.0e-4 & 0.682 & 0.750 & 0.722 & 0.650 & 0.700 & 0.667 & 0.700 & 0.684 & 0.650 & 0.675 & 0.696 & 0.800 & 0.765 & 0.650 & 0.725  \\
8.0e-4 & 0.718 & 0.700 & 0.707 & 0.725 & 0.713 & 0.737 & 0.700 & 0.714 & 0.750 & 0.725 & 0.700 & 0.700 & 0.700 & 0.700 & 0.700  \\
9.0e-4 & 0.714 & 0.750 & 0.737 & 0.700 & 0.725 & 0.714 & 0.750 & 0.737 & 0.700 & 0.725 & 0.714 & 0.750 & 0.737 & 0.700 & 0.725  \\
1.0e-3 & 0.689 & 0.775 & 0.743 & 0.650 & 0.713 & 0.667 & 0.700 & 0.684 & 0.650 & 0.675 & 0.708 & 0.850 & 0.813 & 0.650 & 0.750  \\
1.1e-3 & 0.698 & 0.750 & 0.730 & 0.675 & 0.713 & 0.714 & 0.750 & 0.737 & 0.700 & 0.725 & 0.682 & 0.750 & 0.722 & 0.650 & 0.700  \\
1.2e-3 & 0.707 & 0.725 & 0.718 & 0.700 & 0.713 & 0.700 & 0.700 & 0.700 & 0.700 & 0.700 & 0.714 & 0.750 & 0.737 & 0.700 & 0.725  \\
1.3e-3 & 0.673 & 0.875 & 0.821 & 0.575 & 0.725 & 0.680 & 0.850 & 0.800 & 0.600 & 0.725 & 0.667 & 0.900 & 0.846 & 0.550 & 0.725  \\
1.4e-3 & 0.725 & 0.725 & 0.725 & 0.725 & 0.725 & 0.700 & 0.700 & 0.700 & 0.700 & 0.700 & 0.750 & 0.750 & 0.750 & 0.750 & 0.750  \\
1.5e-3 & 0.711 & 0.675 & 0.690 & 0.725 & 0.700 & 0.722 & 0.650 & 0.682 & 0.750 & 0.700 & 0.700 & 0.700 & 0.700 & 0.700 & 0.700  \\
\bottomrule
\end{tabular}
}
\end{table*}

\begin{table*}[h]
\centering
\caption{Parameter study of DSUDA on different $\alpha$.}
    \label{Ablation_dratio}
    \resizebox{\textwidth}{!}{%
\begin{tabular}{c|ccccc|ccccc|ccccc}
\toprule
\multirow{2}{*}{ $\alpha$} & \multicolumn{5}{c|}{Both Sides}        & \multicolumn{5}{c|}{Left Side}         & \multicolumn{5}{c}{Right Side}         \\\cmidrule{2-16}
                            & NPV   & TNR   & PPV   & TPR   & Acc   & NPV   & TNR   & PPV   & TPR   & Acc   & NPV   & TNR   & PPV   & TPR   & Acc    \\
\midrule
0.01 & 0.673 & 0.875 & 0.821 & 0.575 & 0.725 & 0.654 & 0.850 & 0.786 & 0.550 & 0.700 & 0.692 & 0.900 & 0.857 & 0.600 & 0.750  \\
0.5  & 0.667 & 0.850 & 0.793 & 0.575 & 0.713 & 0.667 & 0.800 & 0.750 & 0.600 & 0.700 & 0.667 & 0.900 & 0.846 & 0.550 & 0.725  \\
0.1  & 0.732 & 0.750 & 0.744 & 0.725 & 0.738 & 0.789 & 0.750 & 0.762 & 0.800 & 0.775 & 0.682 & 0.750 & 0.722 & 0.650 & 0.700  \\
0.3  & 0.689 & 0.775 & 0.743 & 0.650 & 0.713 & 0.667 & 0.700 & 0.684 & 0.650 & 0.675 & 0.708 & 0.850 & 0.813 & 0.650 & 0.750  \\
0.5  & 0.700 & 0.700 & 0.700 & 0.700 & 0.700 & 0.684 & 0.650 & 0.667 & 0.700 & 0.675 & 0.714 & 0.750 & 0.737 & 0.700 & 0.725  \\
0.7  & 0.730 & 0.675 & 0.698 & 0.750 & 0.713 & 0.765 & 0.650 & 0.696 & 0.800 & 0.725 & 0.700 & 0.700 & 0.700 & 0.700 & 0.700  \\
0.9  & 0.682 & 0.750 & 0.722 & 0.650 & 0.700 & 0.682 & 0.750 & 0.722 & 0.650 & 0.700 & 0.682 & 0.750 & 0.722 & 0.650 & 0.700  \\
1    & 0.707 & 0.725 & 0.718 & 0.700 & 0.713 & 0.714 & 0.750 & 0.737 & 0.700 & 0.725 & 0.700 & 0.700 & 0.700 & 0.700 & 0.700  \\
\bottomrule
\end{tabular}
}
\end{table*}

\begin{table*}[h]
\centering
\caption{Parameter study of DSUDA on different $\beta$.}
    \label{Ablation_sae}
    \resizebox{\textwidth}{!}{%
\begin{tabular}{c|ccccc|ccccc|ccccc}
\toprule
\multirow{2}{*}{$\beta$} & \multicolumn{5}{c|}{Both Sides}        & \multicolumn{5}{c|}{Left Side}         & \multicolumn{5}{c}{Right Side}         \\\cmidrule{2-16}
                            & NPV   & TNR   & PPV   & TPR   & Acc   & NPV   & TNR   & PPV   & TPR   & Acc   & NPV   & TNR   & PPV   & TPR   & Acc    \\
\midrule
1.0e-6 & 0.698 & 0.750 & 0.730 & 0.675 & 0.713 & 0.714 & 0.750 & 0.737 & 0.700 & 0.725 & 0.682 & 0.750 & 0.722 & 0.650 & 0.700  \\
1.0e-5 & 0.667 & 0.800 & 0.750 & 0.600 & 0.700 & 0.682 & 0.750 & 0.722 & 0.650 & 0.700 & 0.654 & 0.850 & 0.786 & 0.550 & 0.700   \\
1.0e-4  & 0.707 & 0.725 & 0.718 & 0.700 & 0.713 & 0.737 & 0.700 & 0.714 & 0.750 & 0.725 & 0.682 & 0.750 & 0.722 & 0.650 & 0.700   \\
1.0e-3  & 0.707 & 0.725 & 0.718 & 0.700 & 0.713 & 0.700 & 0.700 & 0.700 & 0.700 & 0.700 & 0.714 & 0.750 & 0.737 & 0.700 & 0.725   \\
1.0e-2  & 0.667 & 0.800 & 0.750 & 0.600 & 0.700 & 0.652 & 0.750 & 0.706 & 0.600 & 0.675 & 0.680 & 0.850 & 0.800 & 0.600 & 0.725  \\
1.0e-1  & 0.643 & 0.675 & 0.658 & 0.625 & 0.650 & 0.650 & 0.650 & 0.650 & 0.650 & 0.650 & 0.636 & 0.700 & 0.667 & 0.600 & 0.650   \\
1.0     & 0.657 & 0.575 & 0.622 & 0.700 & 0.638 & 0.625 & 0.500 & 0.583 & 0.700 & 0.600 & 0.684 & 0.650 & 0.667 & 0.700 & 0.675 \\
\bottomrule
\end{tabular}
}
\end{table*}

\begin{table*}[h]
\centering
\caption{Parameter study of DSUDA on $\eta$.}
    \label{Ablation_sdp}
    \resizebox{\textwidth}{!}{%
\begin{tabular}{c|ccccc|ccccc|ccccc}
\toprule
\multirow{2}{*}{$\eta$} & \multicolumn{5}{c|}{Both Sides}        & \multicolumn{5}{c|}{Left Side}         & \multicolumn{5}{c}{Right Side}         \\\cmidrule{2-16}
                            & NPV   & TNR   & PPV   & TPR   & Acc   & NPV   & TNR   & PPV   & TPR   & Acc   & NPV   & TNR   & PPV   & TPR   & Acc    \\
\midrule
1.0e-6 & 0.681 & 0.800 & 0.758 & 0.625 & 0.713 & 0.714 & 0.750 & 0.737 & 0.700 & 0.725 & 0.654 & 0.850 & 0.786 & 0.550 & 0.700 \\
1.0e-5 & 0.707 & 0.725 & 0.718 & 0.700 & 0.713 & 0.737 & 0.700 & 0.714 & 0.750 & 0.725 & 0.682 & 0.750 & 0.722 & 0.650 & 0.700 \\
1.0e-4  & 0.696 & 0.800 & 0.765 & 0.650 & 0.725 & 0.714 & 0.750 & 0.737 & 0.700 & 0.725 & 0.680 & 0.850 & 0.800 & 0.600 & 0.725 \\
1.0e-3  & 0.711 & 0.675 & 0.690 & 0.725 & 0.700 & 0.765 & 0.650 & 0.696 & 0.800 & 0.725 & 0.667 & 0.700 & 0.684 & 0.650 & 0.675 \\
1.0e-2  & 0.667 & 0.850 & 0.793 & 0.575 & 0.713 & 0.680 & 0.850 & 0.800 & 0.600 & 0.725 & 0.654 & 0.850 & 0.786 & 0.550 & 0.700 \\
1.0e-1  & 0.673 & 0.825 & 0.774 & 0.600 & 0.713 & 0.682 & 0.750 & 0.722 & 0.650 & 0.700 & 0.667 & 0.900 & 0.846 & 0.550 & 0.725 \\
1.0     & 0.750 & 0.675 & 0.705 & 0.775 & 0.725 & 0.765 & 0.650 & 0.696 & 0.800 & 0.725 & 0.737 & 0.700 & 0.714 & 0.750 & 0.725 \\
\bottomrule
\end{tabular}
}
\end{table*}

\subsubsection{ROC Curve Analysis} In Figs.~\ref{ROC_curve} (a-c), we plot the ROC-AUC curves of DSUDA in both-ears, left-ear, and right-ear, respectively. When discriminating the left ear, DSUDA achieved the highest scores in all criteria, i.e., 0/1 classification, Micro- and Macro-average AUC scores of 0.73/0.73, 0.74 and 0.76, respectively. This may be because after aligning the data of the left ear to the data of the right ear, the right ear data may be too aligned and close to each other. Then, the data of the left ear will be easier to distinguish compared to the right ear. In all three settings, our model scores higher on correctly classifying tinnitus subjects than control subjects. This may be due to the more similar patterns of tinnitus patients across different datasets than those of control subjects. It also shows that our model can effectively learn the data patterns of tinnitus patients across datasets.

\subsection{Visualization of Learned Representations}
\subsubsection{Effectiveness of Side-aware Domain Adaptation} As shown in Figs.~\ref{embedding} (a-b), we visualize the raw signals and the representations in the target dataset after domain adaptation, respectively. Each subject contains two signals, i.e., one signal from the left and the right ear, which is the same color. It is obvious that the adapted representations are closer to each other than the raw features, e.g., dark blue and light green. This indicates that our side-aware domain adaptation effectively aligns signals from the left and right sides to a unified domain. SUDA can eliminate the side difference and make it easier for classifiers to detect whether the signals are from the same subject.

\subsubsection{Effectiveness of Disentangled Auto-encoder} We plot the raw signals and the pure features learned by $f_{pe}$ in DSUDA in Figs.~\ref{embedding} (c-d). The tinnitus subjects are colored pink and the control subjects are blue. We can observe that in the raw features, the blue and pink nodes are entangled and hard to distinguish. After disentangling the noise-free classification information, the pink nodes are more clustered together and further away from the blue node. It shows that our model can effectively exclude noise and purify the data.

\section{Discussion}

We set default experimental settings as 1) updating step per batch for DAE and SUDA: 1; 2) learning rates of DAE and SUDA: 1.0e-3; 3) loss ratios $\alpha$, $\beta$, $\eta$: 1.0. We do ablations studies to discuss how the updating steps, learning rates, and loss ratios can affect the performance. Each time, we fix other settings as the default setting, use a fixed random seed, and vary the factors we want to study.

\subsection{Effect of Updating Steps} 
The results of the different updating steps for DAE and SUDA are shown in the tables ~\ref{Ablation_k2} and~\ref{Ablation_k1}, respectively. We can find that the performance of SUDA is stable at different updating steps and shows a slightly increasing trend as the number of updating steps increases. In contrast, the results of varying updating steps of DAE vary widely, and the best results are achieved when the DAE is updated 10 times per batch. The reason may be that the optimization of SUDA can only affect the classification indirectly, while DAE directly extracts features to classify the signal. So minor changes can lead to non-negligible performance differences.

\subsection{Effect of Learning rates} 
We show results with varied learning rates of two model components in Tables~\ref{Ablation_lr2} and \ref{Ablation_lr1}. We can find that the model performance fluctuates up and down as the learning rate increases but is overall stable. Note that we set updating steps as the default value, i.e., 1, thus, according to Table~\ref{Ablation_k2}, the both-ear accuracy would be around 0.725, and the best accuracy (0.775) can not be achieved. This explains why results in ablation studies (Tables~\ref{Ablation_k1}-\ref{Ablation_sdp}) are significantly lower than the best results.

\subsection{Effect of loss ratios}
We summarize results with different loss ratios in Table~\ref{Ablation_dratio}-\ref{Ablation_sdp}. We set the varying ranges based on the values and importance of $L_{d}$ and $L_{side}$: 0.01 to 1 for $\alpha$ ($L_{d}$ ); 1.0e-6 to 1 for $\beta$ and $\eta$ ($L_{side}$). Overall, the performance is stable when changing $\alpha$ and $\eta$, but decreases drastically when $\beta$ is larger than 0.1. The reason lies in that during optimizing DAE, we mainly focusing on representation disentangling, and higher ratios of $L_{side}$ may misguide the optimization direction.

\section{Conclusion and Future Direction}
\subsection{Conclusion}
We propose a novel model, dubbed DSUDA, for EEG-based cross-datasets tinnitus diagnosis. DSUDA integrates disentangled auto-encoder and side-aware unsupervised domain adaptation to adapt the classifier trained on a source domain to the unlabeled target domain. The model can overcome domain discrepancy with domain variance adaptation,  overcome inherent left-right ear pattern difference by subject-level side adaptation, and relieve the negative influence of EEG's low signal-to-noise ratio by disentangling pure information from noises. Our model consistently outperforms state-of-the-art competitors with a large margin regarding classification accuracy and other criteria, demonstrating that the side-aware unsupervised domain adaptation facilitates the model's generalization and the disentangled auto-encoder improves classifying performance.

\subsection{Future Direction}
Our model can be further extended to new scenarios in the future. There are several promising directions. For example, we can apply the DSUDA in other EEG-based tasks, such as activity or emotion recognition. The problems we addressed with our model, i.e., distribution discrepancy and low signal-to-noise ratio,  are universal in EEG-based tasks. Thus, our model can achieve good performance after being modified to fit other tasks. The potential of DSUDA is worth to dig. Or, we can implement DSUDA in real-world clinics or hospitals to assist doctors or therapists in diagnosing and treat tinnitus, which would be of great value to the community and healthcare. Also, we can improve DSUDA with interpretable components to help researchers in medical or bioinformatics reveal latent but essential findings, which may facilitate the cure of tinnitus. Taking other factors into consideration is another direction to improve our model. Although DSUDA already eliminates the influence of ear-side information, many other factors like ages or genders may also change the EEG signal patterns and thus impair classification accuracy.

\section{Acknowledgment}
The authors would like to thank Dr. Hannah Guest and Dr. Roland Schaette for sharing the datasets.

\bibliographystyle{IEEEtran}
\bibliography{bio}

\end{document}